\documentclass[12pt]{article}
\usepackage{graphicx}\usepackage{amssymb}\usepackage{bm}\usepackage{amsfonts}\usepackage{amsmath}\usepackage{cases}
\newcommand{\beq}{\begin{equation}}\newcommand{\eeq}{\end{equation}}\newcommand{\beqa}{\begin{eqnarray}}
\newcommand{\eeqa}{\end{eqnarray}}\newcommand{\w}{\wedge}
\newcommand{\dl}{\bm{\delta}}\newcommand{\nn}{\nonumber}

\newcommand{\h}[1]{\hat{#1}}

\newcommand{\ou}[3]{\underset{#3}{\overset{#1}{#2}}}

\newcommand{\wt}{\widetilde}
\newcommand{\g}{\gamma}
\newcommand{\tg}{\widetilde{\gamma}}

\begin{document}
{\renewcommand{\thefootnote}{\fnsymbol{footnote}}

\begin{center}
{\LARGE  Gravity from a fermionic condensate of a gauge theory}\\
\vspace{1.5em}
Andrew Randono\footnote{e-mail address: {\tt arandono@perimeterinstitute.ca}}
\\
\vspace{0.5em}
The Perimeter Institute for Theoretical Physics \\
31 Caroline Street North\\
Waterloo, ON N2L 2Y5, Canada
\vspace{1.5em}
\end{center}
}

\setcounter{footnote}{0}

\begin{abstract}
The most prominent realization of gravity as a gauge theory similar to the gauge theories of the standard model comes from enlarging the gauge group from the Lorentz group to the de Sitter group. To regain ordinary Einstein-Cartan gravity the symmetry must be broken, which can be accomplished by known quasi-dynamic mechanisms. Motivated by symmetry breaking models in particle physics and condensed matter systems, we propose that the symmetry can naturally be broken by a homogenous and isotropic fermionic condensate of ordinary spinors. We demonstrate that the condensate is compatible with the Einstein-Cartan equations and can be imposed in a fully de Sitter invariant manner. This lends support, and provides a physically realistic mechanism for understanding gravity as a gauge theory with a spontaneously broken local de Sitter symmetry.

\end{abstract}

\section{Introduction}
There are various ways of reformulating and reinterpreting gravity as a gauge theory in direct analogy with the gauge theories of the Standard Model of particle physics \cite{MMoriginal}\cite{Stelle:1979va}\cite{Stelle:1979aj}\cite{Fukuyama:1984}\cite{Percacci:2009ij}\cite{Chamseddine:2010rv}\cite{Lisi:2010td}. In this article we will focus on approaches that enlarge the local gauge group to include all the ingredients of Einstein-Cartan gravity in a single connection. In this framework, the two main ingredients of Einstein-Cartan gravity, the spin connection and the tetrad, are unified into a single connection that takes values in the $(A)dS_4$ Lie algebra. The catch is that the symmetry must be broken in order to regain the Einstein-Cartan equations of motion. This can be accomplished in a minimalist fashion in an explicit symmetry breaking scenario as in the Macdowell-Mansouri mechanism, or through a slightly less minimalistic quasi-dynamic symmetry breaking mechanism. In the latter case, new ingredients must be added to the system which are then constrained in order to break the symmetry. As with all dynamic symmetry breaking mechanisms, the full symmetry is actually retained, but it is realized in a non-linear fashion, and for most practical purposes it is sufficient to fix the gauge that isolates the constrained degrees as frozen constants, effectively ``breaking" the symmetry. 

The main purpose of this article is to improve the quasi-dynamic symmetry breaking mechanism of the Stelle-West model \cite{Stelle:1979aj}\cite{Stelle:1979va}, which we will now briefly review. The model works for both the de Sitter and anti-de Sitter groups, but in this article we will focus on the de Sitter group. $SO(4,1)$, and its double cover, $Spin(4,1)$. The model employs a vector representation, $V^A$ where $A,B,C,...=\{0,1,2,3,4\}$, of $SO(4,1)$ constrained to live in the coset space $SO(4,1)/SO(3,1)$, which is equivalent to constraining the magnitude of the vector field to be a positive constant, $V_A V^A=c^2$. This is accomplished in the action by imposing the constraint via Lagrange multipliers. Given the de Sitter connection $A^{AB}$ and its curvature $F^{AB}$, the Stelle-West action is
\beq
S_{SW}=\sigma \int_M V^A \,\epsilon_{ABCDE} \,F^{BC}\w F^{DE} +\alpha \left( c^2- V_A V^A\right).
\eeq
Varying with respect to the Lagrange mutiplier $\alpha$ imposes the constraint. To see that this is equivalent to Einstein-Cartan gravity, at least at the local level, on the constraint surface we can always choose a gauge where $V^A=(0,0,0,0,c)$. In this frame, if we identify the tetrad as
\beqa
-\ell\, DV^I= -\ell c\,A^I{}_4=e^I \quad\quad \quad DV^4=0 
\eeqa
where $\ell$ is related to the cosmological constant by $\ell=\sqrt{(3/|\Lambda|)}$, we regain the Einstein-Cartan action together with a topological term:
\beqa
S_{SW}&\stackrel{|V|=c}{=}&-\frac{3}{64\pi G\Lambda}\int_M \epsilon_{IJKL}\,R^{IJ} \w R^{KL}\nn\\
& & +\frac{1}{32 \pi G}\int_M \epsilon_{IJKL}\,e^I \w e^J \w R^{KL} -\frac{\Lambda}{6}\, \epsilon_{IJKL}\,e^I\w e^J\w e^K\w e^L \label{SWReducedAction}
\eeqa
provided $c\sigma=-\frac{3}{64\pi G \Lambda }$. The remaining equations of motion obtain from varying the action with respect to $V^A$ simply constrain the Lagrange multiplier to be proportional to the square of the Weyl tensor, and they do not constrain the field equations beyond this. 

The deficiencies of this model come from the use of the five-dimensional vector field $V^A$ as the order parameter to break the symmetry. Such an object has little in common with the familiar ingredients of gravity and the standard model of particle physics, so in this sense the mechanism is somewhat unphysical. Furthermore, the vector field does not itself have an explicit kinetic term associated with it. Since the constraint $V^A V_A=c^2$ is imposed by hand via the use of Lagrange multipliers in the action we will refer to the mechanism as quasi-dynamic. Preferably, such a constraint would follow from a true dynamic symmetry breaking wherein the field relaxes to an energetically or otherwise preferred minimum of a realistic potential, thereby attaining the constraint dynamically. Being such an unfamiliar object, it is difficult to build physically motivated kinetic terms an potentials for the vector field $V^A$.

The purpose of this paper is to address some of these deficiencies. The main goal of the present work is rather modest -- rather than searching directly for a full dynamic symmetry breaking mechanism, we will simply introduce a new model where the {\it ingredients} used in the construction are more amenable to this ultimate goal. More specifically, we will show that one can replace the vector field $V^A$ with a multiplet of ordinary Dirac spinor fields to serve as the order parameters to break the symmetry. Motivated by symmetry breaking models in condensed matter physics, such as the BCS model of superconductivity, the symmetry in our model will be broken by a highly symmetric fermionic condensate. For other BCS motivated effects of fermionic condensation in cosmology see \cite{Alexander:2009uu}. We will implement the condensate quasi-dynamically via Lagrange multipliers in the action, and show that the resulting field equations are internally consistent and locally equivalent to Einstein-Cartan gravity. However, as in most symmetry breaking scenarios, this type of mechanism can give rise to new, global degrees of freedom of a topological nature, which we will discuss at the end of the article. Since the new model uses ingredients that are closer to those of well-understood and tested theories, the hope is that the model can be extended to a genuinely dynamical mechanism for breaking the symmetry of a de Sitter gauge theory to the Lorentz group to obtain Einstein-Cartan gravity.

\section{The model}
As mentioned in the introduction, the symmetry breaking model we will present replaces the vector field $V^A$ of the Stelle-West model with a condensate of a fermionic multiplet, $\psi_i$. Our model shares some elements in common with a model for supergravity presented in \cite{Tung:1995cj}. The basis of the mechanism we will construct comes from the recognition that Dirac spinors, although normally viewed as representations of the double cover of the Lorentz group, can be seen as representations of a larger group. For disambiguation, we refer to Dirac spinors at their most primitive level simply as 4-component complex (anti-commuting) vectors subject to the inner product 
\beq
\langle \phi |\psi \rangle_D =\bar{\phi} \psi=\phi^\dagger \gamma^0 \psi \,.
\eeq
As shown in Appendix \ref{AppendixA}, the simply connected group that preserves this inner product is $Spin(4,2)$, the double cover of the conformal group in four dimensions. This group has two natural subgroups, the anti-de Sitter and de Sitter groups, $Spin(3,2)$ and $Spin(4,1)$ respectively, which can be distinguished by imposing that the group must also preserve an additional inner product (see Appendix \ref{AppendixA}). 
For definiteness, for the remainder of the paper we will work with the de Sitter group, since it is observationally preferred and because the structure of the group allows for interesting geometric structures in the context of the de Sitter gauge theory of gravity, which we will discuss at the end of the paper. The salient point is that Dirac spinors naturally form a representation of the de Sitter group $Spin(4,1)$.

We now begin with the construction of the symmetry breaking model. Consider a principle G-bundle where the gauge group is the de Sitter group, $G=Spin(4,1)$. We begin with a multiplet of $N$ spinor fields $\psi_{i}$ where $i=1,2,...N$, and $N$ has yet to be determined. Consider the following action\footnote{Since we are working with $-,+,+,+$ signature, we define $\bar{\psi}=\psi^{\dagger}\gamma^0$ so that $(\bar{\psi}\psi)^{\dagger}=\bar{\psi}\psi$ if the components of $\psi$ are Grassman numbers.}:
\beq
S=\sigma \int_M DD\bar{\psi}_i\w DD\psi_i \,.
\eeq
Here it is understood that the exterior covariant derivative $D=d+A$ is the covariant derivative of the de Sitter connection and we are summing over repeated indices of the multiplet. Using $DD\psi=F\psi$ the action can be written
\beqa
S&=&-\sigma \int_M \bar{\psi}_i F\w F\psi_i \nn\\
&=& \sigma \int_M Tr \left(\psi_i\bar{\psi}_i\,F\w F\right)
\eeqa
In the last term, the object $\psi_i \bar{\psi}_i$ is recognized as a complex $4\times 4$ matrix. Since the Clifford algebra $Cl_{3,1}$ forms a basis for the set of complex $4\times 4$ matrices, we can expand $\psi_i \bar{\psi}_i$ as a linear combination of elements of the Clifford algebra. A convenient basis is given by $\Gamma^{\hat{A}}=\{\frac{1}{2}, \frac{1}{2} i\gamma^I, \frac{1}{2}\star \!\gamma^I ,\frac{i}{2}\gamma^{[I}\gamma^{J]}, \frac{1}{2}\star\}$ where $\hat{A}$ is an index in the sixteen dimensional Clifford algebra and $\star \equiv -i\gamma_5$. The normalization has been chosen so that each distinct element has norm $+1$ or $-1$ under the trace as an inner product, and each element in the basis satisfies $(\Gamma^{\hat{A}})^{\dagger}=-\gamma^0 \Gamma^{\hat{A}} \gamma^0$, the reason for which will become clear shortly. For convenience, it is useful to denote the fermion multiplet $\Psi$ defined by 
\beq
\Psi=\left[\begin{matrix}\psi_1 \\ \psi_2 \\ \vdots \\ \psi_N\end{matrix} \right] \quad \quad \bar{\Psi}M\Psi = \bar{\psi}_1 M\psi_1+\bar{\psi}_2 M \psi_2 +\cdots +\bar{\psi}_N M \psi_N
\eeq
where it is understood that we will use the short-hand notation in the multiplet representation (unless otherwise specified)
\beq
M \Psi= \begin{bmatrix}
  M      & \cdots & 0      \\
  \vdots & \ddots & \vdots \\ 
  0      & \cdots & M 
\end{bmatrix}\left[\begin{matrix}\psi_1 \\ \vdots \\ \psi_N\end{matrix} \right]
\eeq
The expansion of the matrix in the Clifford algebra, sometimes referred to as a Fierz decomposition (see e.g. \cite{Ortin}), is then given by 
\beqa
\psi_i\bar{\psi}_i&=&\sum_{\hat{A}} \bar{\Psi} \Gamma_{\hat{A}}\Psi \,\Gamma^{\hat{A}}\\
&=& -\frac{1}{4}\left(\bar{\Psi}{\Psi}\,1 -\bar{\Psi}\!\star\!\Psi\,\star-i\bar{\Psi}\gamma_I{\Psi}\,i\gamma^I+\bar{\Psi}\star \!\gamma_I {\Psi}\,\star\!\gamma^I+\frac{1}{2}\,i\bar{\Psi}\gamma_{[I}\gamma_{J]}{\Psi}\,i\gamma^{[I}\gamma^{J]}\right)\,. \nn\label{BilinearExpansion}
\eeqa
We note that the matrix $\psi_i\bar{\psi}_i$ satisfies the adjoint condition $(\psi_i\bar{\psi}_i)^{\dagger}=-\g^0 (\psi_i\bar{\psi}_i)\g^0$. An arbitrary complex $4\times 4$ matrix, $M$ satisfying $M^{\dagger}=-\g^0 M\g^0$ has $16$ real degrees of freedom. Since a single spinor field $\psi$ has only 8 real degrees of freedom, it is now clear why we have to work with a spinor multiplet. It may be sufficient two work with a doublet, since the doublet itself has 16 real degrees of freedom. However, we are always safe with a spinor quadruplet, since one can always find a set of four spinors, $\psi_i$, such that $M=\psi_i \bar{\psi}_i$. To see this, we first note that any matrix $M$ satisfying $M^{\dagger}=-\g^0 M\g^0$ can be written $M=-A \,i\g^0$ where $A$ is Hermitian. All Hermitian matrices can be expanded in a basis of orthonormal eigenvectors $\phi_i$ with eigenvalues $\alpha_i$ so that $A=i\alpha_i \phi_i \phi^\dagger _i$, where the extra factor of $i$ is necessary if $\phi_{i}$ are to be anti-commuting Grassman variables. Defining $\psi_i = \sqrt{\alpha_i} \phi_i$, we then have $M=\psi_i \bar{\psi}_i$. Thus, for the remainder of the paper we will assume $N=4$.

The symmetry is broken when the spinor fields form a condensate and the matrix $\psi_i \bar{\psi}_i$ reduces to the vacuum expectation value $\psi_i \bar{\psi}_i\rightarrow\langle \psi_i \bar{\psi}_i\rangle$. The VEV (vacuum expectation value) is chosen to break the $Spin(4,1)$ symmetry to $Spin(3,1)$. The only assumption we will make, initially, is that the condensate that forms is both {\it isotropic} and {\it homogenous} with respect to the subgroup $Spin(3,1)$, which acts as a stabilizer for the VEV. Isotropic means that the VEV does not distinguish a preferred direction at any point, equivalent to the assumptions
\beq
\langle\bar{\Psi}\gamma^I \Psi \rangle=0\quad \quad \langle\bar{\Psi}\gamma_5 \gamma^I  \Psi \rangle=0 \quad \quad \langle\bar{\Psi}\gamma^{[I}\gamma^{J]} \Psi \rangle=0\,. \label{VEVisotropy}
\eeq
Homogenous means that the remaining two degrees of freedom are constant:
\beq
\langle\bar{\Psi}\!\star\!\Psi \rangle=-c_1=constant \quad\quad \langle\bar{\Psi} \Psi \rangle=c_2=constant \label{VEVhomogeneity}
\eeq
where we have defined $\star=-i\gamma_5$ so that the constant is real. It is worth mentioning that had we began with only one spinor then both constants above would be zero. With four spinors in the multiplet, we have sufficient number of degrees of freedom to obtain a non-trivial result. In total, the condensate that forms satisfies
\beq
\langle \psi_i \bar{\psi}_i \rangle =\frac{1}{4} \left(\langle\bar{\Psi}\!\star\! \Psi \rangle\star - \langle \bar{\Psi} \Psi \rangle\right)=-\frac{1}{4}\left(c_1 \star +c_2\right)
\eeq
Inserting this into the action and defining the constants
\beq
\frac{\theta}{8\pi^2} =- \frac{\sigma \langle\bar{\Psi} \Psi \rangle}{4} \quad \quad \frac{1}{\kappa}=\frac{\sigma\langle\bar{\Psi}\star \Psi \rangle}{4}
\eeq
we obtain
\beq
S=\frac{1}{\kappa}\int_M Tr\left(\star F\w F\right) +\frac{\theta}{8\pi^2}\int_M Tr\left(F\w F\right)
\,. \label{MMAction}
\eeq
This is the Macdowell-Mansouri action \cite{MMoriginal} with a topological $\theta$-term provided $\frac{1}{\kappa}=-\frac{3}{16\pi G\Lambda}$. To see that this is equivalent to (\ref{SWReducedAction}) up to topological terms, we compute the trace over the Clifford algebra explicitly:
\beqa
S&=&\frac{1}{\kappa}\int_M \star R\,R -\frac{2}{\ell^2} \star e\,e\,R +\frac{1}{\ell^4}\star e\,e\,e\,e 
+\frac{\theta}{8\pi^2}\int_M R\,R -\frac{1}{\ell^2} \left(T\,T +2 e\,e\,R\right) \nn\\
&=& \frac{1}{32\pi G}\int_M  \epsilon_{IJKL}\,e^I\,e^J\,R^{KL} -\frac{\Lambda}{6}\,\epsilon_{IJKL}\,e^I\,e^J\,e^K\,e^L \\
&& +\int_M \frac{\theta}{8\pi^2} \left(e_I\,e_J \,R^{IJ}-T_I\,T^I \right) -\frac{\theta}{16\pi^2} R_{IJ}\,R^{IJ}-\frac{3}{64 \pi G\Lambda} \epsilon_{IJKL}\,R^{IJ}\,R^{KL}\nn
\eeqa

The three terms in the last line are topological terms, respectively proportional to the Nieh-Yan class, the second Chern class, and the Euler class. As proposed in \cite{RandonoMercuri}, the theta-parameter can be directly related to the Immirzi parameter of Loop Quantum Gravity, and it can be argued to arise naturally from arguments involving quantum mechanical tunneling between degenerate vacua.

\subsection{Constraining the theory to obtain a VEV}
As we have seen, if the fermion field can obtain a VEV that is isotropic and homogenous with respect to the induced stabilizer subgroup, then we regain Einstein-Cartan gravity with a topological term. It has yet to be shown that this can be implemented consistently, which we will now show. It is essential, and not guaranteed, that the constraints imposed are compatible with the Einstein-Cartan equations of motion, and that they can be implemented in a $Spin(4,1)$ invariant manner, which is necessary if the model has any potential for extension to a genuine dynamical symmetry breaking mechanism. Since the ultimate goal is construct a dynamical theory wherein the field naturally relaxes to the given vacuum expectation value, as a first step to achieving this goal we will now demonstrate how the VEV can be implemented in the action by employing Lagrange multipliers to impose a set of constraints. In order to remain faithful to the full underlying $Spin(4,1)$ symmetry, we will seek to define the VEV in a gauge invariant manner, as opposed to defining it by the conditions (\ref{VEVisotropy}) and (\ref{VEVhomogeneity}), which are valid only in a particular gauge. 

To express the constraints in a $Spin(4,1)$ invariant manner, we first reformulate the expansion (\ref{BilinearExpansion}) in a form more suitable to a $Spin(4,1)$ description. The form of (\ref{BilinearExpansion}) is particularly convenient in the context of the $Spin(3,1)$ phase, since each term in the expansion lies in a subspace of the Clifford algebra that is closed under $Spin(3,1)$ transformations. That is, scalars, psuedo-scalars, vectors, axial-vectors and bivectors remain scalars, psuedo-scalars, vectors, axial-vectors and bivectors under a $Spin(3,1)$ transformation. However, a generic $Spin(4,1)$ transformation will not preserve these subspaces. But, one can combine the vector and psuedo-scalar, and the axial-vector and bivector subspaces into two spaces that {\it are} closed under $Spin(4,1)$. To see this, it is convenient to embed the Clifford algebra $Cl_{3,1}$ into the higher dimensional $Cl_{4,1}$ as follows. Define $\tg^A$, where $\{A,B,C,...\}$ take values in $\{0,1,2,3,4\}$, by
\beq
\tg^I=\g^I \quad \quad \tg^4=\g_5\,.
\eeq
The new algebra satisfies
\beq
\tg^A\tg^B +\tg^B\tg^A=2\eta^{AB}
\eeq
where $\eta^{AB}=diag(-1,1,1,1,1)$, so products of these elements form the Clifford algebra $Cl_{4,1}$ in the usual way. Consider now the bilinears $i\bar{\Psi}\tg^A\Psi$ and $i\bar{\Psi}\tg^{[A}\tg^{B]}\Psi$, with the $i$ in front so that the components are real. The components are 

\beqa
i\bar{\Psi}\tg^I\Psi=i\bar{\Psi}\g^I\Psi &\quad \quad& i\bar{\Psi}\tg^4\Psi=i\bar{\Psi}\g_5\Psi  \nn\\
i\bar{\Psi}\tg^{[I}\tg^{J]}\Psi =i\bar{\Psi}\g^{[I}\g^{J]}\Psi &\quad \quad&  i\bar{\Psi}\tg^{[4}\tg^{J]}\Psi =i\bar{\Psi}\g_5\g^{J}\Psi\,.
\eeqa
Now, $\frac{1}{2}\tg^{[A}\tg^{B]}$ itself furnishes a representation of $\mathfrak{spin}(4,1)$, as discussed further in Appendix \ref{AppendixA}. Furthermore, it can easily be seen that it preserves the respective scalar, vector, and bivector spaces in which the bilinears $\bar{\Psi}\Psi$, $i\bar{\Psi}\tg^A\Psi$, and $i\bar{\Psi}\tg^{[A}\tg^{B]}\Psi$ reside. Thus, the matrix $\psi_i \bar{\psi}_i$ can now be decomposed into this $Spin(4,1)$ irreducible basis to give 
\beqa
\psi_i\bar{\psi}_i&=& -\frac{1}{4}\left(\bar{\Psi}{\Psi}\,1 -i\bar{\Psi}\tg_A{\Psi}\,i\tg^A+\frac{1}{2}\,i\bar{\Psi}\tg_{[A}\tg_{B]}{\Psi}\,i\tg^{[A}\tg^{B]}\right)\,. \label{BilinearExpansion}
\eeqa

We can now state clearly the constraints that reduce the spinor field to its VEV as constraints on the magnitude of these objects.  In total the constraints can be written in a gauge covariant way as follows:
\beq
i\bar{\Psi}\tg_A\Psi \,i\bar{\Psi}\tg^A\Psi =c_1^2 \quad\quad \bar{\Psi}\Psi\,\bar{\Psi}\Psi =c_2^2  \quad\quad \bar{\Psi}\tg^{[A}\tg^{B]}\Psi=0\label{VEV2}
\eeq
where $c_1$ and $c_2$ are both real numbers (implying $\bar{\Psi}\tg_A\Psi$ is spacelike). Thus we have $c_2^2 =\langle \bar{\Psi}\Psi\rangle^2$, and we can use the $Spin(4,1)$ gauge freedom to choose a gauge where $c_1^2=\langle \bar{\Psi}\!\star\! \Psi \rangle ^2$. 

Since these constraints are manifestly $Spin(4,1)_M$ covariant, they can be implemented through Lagrange multipliers in a $Spin(4,1)_M$ invariant action. Thus,  to impose these constraints we introduce two scalar-valued four-form Lagrange multipliers $\alpha$ and $\beta$, and one bivector valued four-form $\gamma_{AB}=\gamma_{[AB]}$. The total, $Spin(4,1)_M$ invariant action is
\beqa
S=\sigma \int_{M} DD\bar{\Psi}\w DD\Psi &+& \alpha \left(c_1^2 - i\bar{\Psi} \tg_{A} \Psi \,i\bar{\Psi} \tg^A \Psi\right) \nn\\
&&+\beta \left(c_2^2 - \bar{\Psi} \Psi \,\bar{\Psi} \Psi\right) +\gamma_{AB} \,\bar{\Psi} \tg^{[A}\tg^{B]}\Psi \,.
\eeqa
Varying with respect to the Lagrange multipliers clearly implements the constraints (\ref{VEV2}), which can then be inserted back into the action to yield the Macdowell-Mansouri form of the Einstein-Cartan action, (\ref{MMAction}). The field equations obtained from $\dl A$ on the condensate then are locally equivalent to the Einstein-Cartan equations of motion. On the other hand, we still have to check that varying with respect to $\psi_i$ or $\bar{\psi}_i$ does not give additional constraints that will further constrain the equations of motion. In fact it does not -- the fixed points of the action upon varying these fields can be solved algebraically to fix the Lagrange multipliers and nothing more, as we will now show. 

Consider then the variation of the action with respect to $\bar{\psi}_i$. The equations of motion obtained by the fixed point are
\beqa
F\,F\,\psi_i=-2i\alpha\,J\psi_i -2\beta \,\bar{\Psi}\Psi \,\psi_i +\g_{AB}\,\tg^A \tg^B\psi_i\,.\label{NewEOM}
\eeqa
where we have defined $J\equiv J^A\tg_A =i\bar{\Psi}\tg^A \Psi \,\tg_A$.
We proceed by multiplying both sides by $\bar{\psi}_i$ from the right to obtain the matrix $\psi_i \bar{\psi}_i$. On the constraint surface imposed by the variation with respect to the Lagrange multipliers, one can chose a gauge where $\psi_i \bar{\psi}_i=-\frac{1}{4}\left(c_1 \star +c_2\right)$. From this form it is clear that as long as either $c_1$ or $c_2$ is non-zero, the matrix is invertible, and this property will hold in any gauge. Thus, inverting this matrix we have
\beqa
F\,F=-2i\alpha\,J -2\beta \,\bar{\Psi}\Psi +\g_{AB}\,\tg^A \tg^B\,.
\eeqa
Using the trace properties of the Clifford algebra, this equation can be solved for the Lagrange multipliers to give
\beqa
\alpha\ &=&-\frac{1}{8i \,J\cdot J} \ Tr\left(J F\w F \right)\nn\\
\beta \ &=& -\frac{1}{8\bar{\Psi}\Psi} \ Tr \left(F\w F \right) \nn\\
\gamma^{AB}&=& 0 \,.
\eeqa
Inserting these expressions back into (\ref{NewEOM}) we obtain
\beqa
F\w F \psi_i=\frac{1}{4 J\cdot J} \ Tr(J F\w F) \,J\psi_i +\frac{1}{4} \,Tr(F\w F) \psi_i \,.
\eeqa
This equation reveals an (apparent) additional constraint on the curvature of the form
\beq
Tr(\tg^A F\w F)=\frac{J^A}{J\cdot J}\ Tr(J F\w F)
\eeq
which can be written in terms of a perpendicular projection operator as follows
\beq
\left(\delta^A{}_B -\frac{J^A J_B}{J\cdot J} \right) \ Tr(\tg^B F\w F)=0 \label{Projection}\,.
\eeq
The expression above says that the components of $Tr(\tg^A F\w F)$ perpendicular to $J^A$ are constrained to be zero, whereas the component parallel to $J^A$ is left free. At first sight this appears to be an additional constraint on the curvature, however, the constraint already holds identically on the equations of motion obtained by varying the action with respect to $\dl A$. To see this, choose a gauge where the current $J^A=(0,0,0,0,c_1)$, or $J^I=0$ and $J^4=c_1$. In this gauge, the component of $Tr(\tg^A F\w F)$ perpendicular to $J^A$ is proportional to $Tr(\star F\w F)$, which necessarily must be a free variable since this expression evaluated on the Einstein-Cartan equations of motion is proportional to the square of the Weyl tensor. The remaining components in this gauge are
\beq
Tr(\gamma^I F\w F)\ \sim\  \epsilon^I{}_{JKL}\,T^J \w \left(R^{KL}-\frac{1}{\ell^2}e^K \w e^L\right)\,. 
\eeq
Recall that in this gauge the equations of motion obtained from the variation $\dl A$ on the condensate are precisely the Einstein-Cartan equations of motion (we are simply trying to determine if there are additional constraints on the curvature due to the condensate), which are given by
\beqa
\epsilon^I{}_{JKL}\,e^J \w \left(R^{KL}-\frac{1}{\ell^2}e^K \w e^L\right)&=& 0 \nn\\
D\left( e^I \w e^J \right) &=& 0\,.
\eeqa
From this it is a simple matter to show that
\beq
Tr(\gamma^I F\w F) \approx 0
\eeq
where the weak equality holds on the fixed points of the action obtained by the variation $\dl A$. We note that this holds even when it is not assumed that the tetrad is invertible. In total then, the apparent additional constraint (\ref{Projection}) already holds on the Einstein-Cartan equations of motion and does not constrain the field equations further. We can, therefore, insert (\ref{Projection}) back into (\ref{NewEOM}) to obtain
\beqa
F\w F \,\psi_i=\left(\frac{1}{4} \ Tr(\tg^A F\w F)\tg_A+\frac{1}{4} \,Tr(F\w F) \right)\psi_i \,.
\eeqa
which we now recognize as a trivial tautology since the object in brackets on the right hand side is nothing more than the expansion of $F\w F$ in a Clifford algebra basis.

We conclude that the equations of motion obtained from the variations $(\dl A^{AB} , $ $\dl \psi_i,\dl\bar{\psi}_j , \dl \alpha ,\dl \beta , \dl \gamma^{AB}) $ are precisely a $Spin(4,1)$ invariant version of the Einstein-Cartan equations of motion together with a complete specification of the Lagrange multpliers in terms of the other fields, {\it and nothing more}. We offer this as strong evidence that the Einstein-Cartan equations of motion can consistently be regarded as the symmetry broken phase of a de Sitter gauge theory obtained by a homogenous and isotropic fermion condensate.

\section{Global features of the condensate}
As with any symmetry breaking mechanism, there remains the possibility that the vacuum obtained by the order parameter relaxing to its vacuum expectation value is not unique, but allows for a class of locally identical but globally distinct condensates. In fact, as we will now show, this mechanism allows for an infinite class of physically distinct vacua that differ by global topological properties of the condensate. This was explored in the context of explicit symmetry breaking in \cite{RandonodSSpaces}, but the results of that paper can be extended to our quasi-dynamic symmetry breaking mechanism allowing for a new interpretation of the class of states. The main observations of this section are not specific to our model, but also hold for the Stelle-West model. However, since the vector field $J^A$ in our model is directly related to the familiar pseudo-scalar and vector current of the spinor field, the expressions assume a more physical and intuitive character in terms of the fermionic condensate.

To construct the class of degenerate states we first need to relax our notion of isotropy and homogeneity of the condensate as a global property, to simply demand that these properties hold in a local neighborhood. More specifically, let us fix the topology of the manifold to be that of de Sitter space, $M=\mathbb{R}\times \mathbb{S}^3$. Given any open neighborhood $\mathcal{U}$ isomorphic to $\mathbb{R}^4$ and embedded in $M$, the condensate is said to be {\it locally} homogenous and isotropic if there exists a de Sitter gauge such that in the region $\mathcal{U}$ the condensate satisfies (dropping the $\langle \cdot \rangle$ notation for the VEV)
\beqa
\bar{\Psi}\gamma^I \Psi =0\quad \quad \bar{\Psi}\gamma_5 \gamma^I  \Psi =0 \quad \quad \bar{\Psi}\gamma^{[I}\gamma^{J]} \Psi =0 \nn\\
\bar{\Psi}\!\star\!\Psi =c_1 \quad\quad  \bar{\Psi} \Psi =c_2\quad \quad\,.
\eeqa
In fact this weaker condition can be summarized by the constraints that we have already imposed in our model, namely
\beq
i\bar{\Psi} \tg_A \Psi \, i\bar{\Psi}\tg^A \Psi=c_1^2 \quad\quad (\bar{\Psi}\Psi )^2=c_2^2 \quad\quad \bar{\Psi}\tg^{[A}\tg^{B]} \Psi =0
\eeq
since the first of these constraints only demands that there exists a gauge where $\bar{\Psi}\star \Psi =c_1$. The weaker, local definition of the condensate allows for global topological degrees of freedom that will distinguish a class of physically distinct vacua. 

To see this, let us first recall the construction of an infinite class of globally distinct but locally de Sitter solutions presented in \cite{RandonodSSpaces}, which we will then reconstruct in the context of our quasi-dynamic symmetry breaking mechanism. The construction exploits a residual symmetry of a subsector of Einstein-Cartan theory that can be interpreted as a remnant of a larger gauge symmetry upon symmetry breaking. Given a solution $A=\omega+\frac{1}{\ell}\gamma_5 e$ to the Einstein-Cartan field equations, generically the gauge transformed field ${}^gA=gAg^{-1}-dg \,g^{-1}$ will not be a solution if $g$ is an arbitrary element of the de Sitter group. If however, we restrict ourselves to a flat ($F=0$) solution, the induced curvature ${}^g F=gFg^{-1}$ is clearly still flat. There is a unique, distinguished flat solution where the tetrad is globally non-degenerate, namely de Sitter space itself. Given a set of generating elements of $\pi_3(Spin(4,1))=\mathbb{Z}\oplus\mathbb{Z}$ denoted by $\ou{m}{g}{n}$ with $m$ and $n$ integers, one can construct an infinite class of new solutions to the field equations beginning with the de Sitter solution $\ou{0}{A}{0}$ by gauge transforming the field $\ou{0}{A}{0}\rightarrow \ou{m}{A}{n}=\ou{m}{g}{n} \ou{0}{A}{0}\ou{m}{g}{n}{}^{-1}-d\ou{m}{g}{n}\,\ou{m}{g}{n}{}^{-1}$ and extracting the new tetrad $\ou{m}{e}{n}$ and spin connection $\ou{m}{\omega}{n}$. The new solutions are automatically guaranteed to be flat, and are therefore solutions to the Einstein-Cartan field equations. However, whereas the de Sitter solution $\ou{0}{e}{0}$ is non-denegerate, each of the solutions $\ou{m}{e}{n}$ for $m-n\neq 0$ has degeneracies on well behaved surfaces. Although the solutions are gauge related with respect to the full de Sitter group, $Spin(4,1)$, the symmetries of the solutions space of Einstein-Cartan gravity are restricted to local $Spin(3,1)$ transformations (together with diffeomorphisms), and the solutions are physically distinct with respect to this limited set of symmetries. In fact, one can construct a geometric charge invariant under $Spin(3,1)_M\rtimes Diff_4(M)$ distinguishing these solutions. This is accomplished by first identifiying a distinguished hypersurface, $\Sigma_0$ in a diffeomorphism invariant as the surface which is preserved under time reversal symmetry, i.e. the ``throat" of the de Sitter solution. The three volume of this surface can be calculated to give
\beq
{}^3 \ou{m}{V}{n}(\Sigma_0)=\int_{\Sigma_0} \frac{1}{3!}n^I \epsilon_{IJKL}\ou{m}{e}{n}{}^J \w \ou{m}{e}{n}{}^K \w \ou{m}{e}{n}{}^L=2\pi^2 \ell^3 (1-q) \label{mnVolume}
\eeq
 where $q\equiv m-n$ and $n^I$ is the lift of the normal to the surface. Our task is now to reconstruct these solutions in the context of the quasi-dynamic symmetry breaking mechanism to see if this class of solutions can still exist in a $Spin(4,1)$ invariant theory.
 
 \subsection{Explicit versus dynamic symmetry breaking}
 The key property of the de Sitter algebra that allows for the separation of the tetrad from the spin connection is the identification of a reductive splitting of the algebra into a stabilizer subgroup plus the transvections (see \cite{Wise:MMGravity}\cite{Wise:2009fu}), such that the reductive split has the property of being {\it symmetric}. This term refers to the existence of an involution operation under which the Lie algebra is graded with the elements of the stabilizer being the even elements and the transvections (translations) being the odd elements. More specifically, given a reductive split of Lie algebra $\mathfrak{g}=\mathfrak{h}\oplus \mathfrak{p}$, and an arbitrary elements $\alpha=\lambda+\eta$ where $\lambda\in \mathfrak{h}$ and $\eta \in \mathfrak{p}$, an involution is an operation such that
\beq
\alpha=\lambda+\eta \quad \longrightarrow \quad \alpha_{\star}=\lambda -\eta\,.
\eeq

The difference between the (quasi-) dynamic versus explicit symmetry breaking can be characterized succinctly from properties of the involution. In the explicit symmetry breaking scenario, the involution is assumed to be a fixed operation independent of the gauge, whereas in the dynamic situation, the involution is gauge covariant. For example, suppose $\mathfrak{g}=\mathfrak{spin}(4,1)$ and $\mathfrak{h}=\mathfrak{spin}(3,1)$. In the Clifford algebra representation given above, the explicit symmetry breaking scenario fixes the involution to be
\beq
\alpha_\star=\gamma_5 \,\alpha \,\gamma_5\,.
\eeq
In the dynamic symmetry breaking mechanism the involution is itself constructed out of dynamic, gauge covariant fields living in the coset space $G/H=Spin(4,1)/Spin(3,1)$, which itself can be identified as a geometry isomorphic to de Sitter space (for a discussion of related issues in $2+1$ gravity, see \cite{Percacci:1986hu}). In our case, defining $\widetilde{J}=\widetilde{J}^A \, \tg_{A}=\tg_A \,J^A/|J|$, the involution is given by
\beq
\alpha_\star =\widetilde{J} \,\alpha \,\widetilde{J}\,.
\eeq
The tetrad can then be related to the non-zero components of $D\widetilde{J}^A$. For example, in the canonical gauge where $\widetilde{J}^A=(0,0,0,0,1)$, we have\footnote{The symbol $\stackrel{*}{=}$ is simply an indicator that the equality holds only in a particular gauge.}
\beq
D\widetilde{J}^4\stackrel{*}{=}0 \quad \quad e^I\stackrel{*}{=}-\ell \,D\widetilde{J}^I \,.
\eeq

This allows us to write down expressions such as (\ref{mnVolume}) in a gauge covariant manner. Furthermore, the interpretation of the expression takes on different character in various choices of the gauge. Whereas in the explcit symmetry breaking scenario (\ref{mnVolume}) is restricted to be a $Spin(3,1)_M \rtimes Diff_4(M)$ invariant expression, in the dynamic scenario, the expression can be written (dropping the $m$ and $n$ labels for convenience)
\beq
\frac{{}^3 \ou{m}{V}{n}(\Sigma_0)}{2\pi^2 \ell^3}= \frac{1}{2\pi^2}\int_{\Sigma_0}  \frac{1}{3!}\epsilon_{ABCDE} \,n^A\, \wt{J}^B\,D\wt{J}^C \w D\wt{J}^D \w D\wt{J}^E
\eeq
where $n^A$ is a timelike unit vector such that $n_A n^A=-1$, $n_B \wt{J}^B=0$, and the pull-back of $n_A D\wt{J}^A$ to $\Sigma_0$ is zero. In the canonical gauge where $\widetilde{J}^A=(0,0,0,0,1)$, the expression is trivially equivalent to (\ref{mnVolume}), and can therefore be related to the difference of two Chern-Simons functionals. However, it is also convenient to choose a gauge where $n^A=(1,0,0,0,0)$ and use the remaining $Spin(4)$ gauge freedom to gauge away the connection coefficients ($A=0$) on $\Sigma_0$, which can always be done since $\Sigma_0\simeq \mathbb{S}^3$ and the three sphere is parallelizable. In this gauge we have $\wt{J}^A=(0,\wt{J}^{\h{a}})$, since $n_A \wt{J}^A=0$, where the hatted indices take values $\{\h{a},\h{b},\dots\}=\{1,2,3,4\}$. The expression then becomes 
\beq
\frac{{}^3 \ou{m}{V}{n}(\Sigma_0)}{2\pi^2 \ell^3} \stackrel{*}{=} \frac{1}{2\pi^2}\int_{\Sigma_0}  \frac{1}{3!}\epsilon_{\h{a}\h{b}\h{c}\h{d}} \, \wt{J}^{\h{a}}\,d\wt{J}^{\h{b}} \w d\wt{J}^{\h{c}} \w d\wt{J}^{\h{e}}\,. \label{mnVolume2}
\eeq
The right hand side is a well-known expression for the index or winding number of the vector field, which can now be viewed as a map $\wt{J}^{\h{a}}:\mathbb{S}^3 \rightarrow \mathbb{S}^3$. Recalling that the integral above is equal to $1-q$, this result says, for example, that the winding number of $\wt{J}^{\h{a}}$ for de Sitter space in the trivial gauge is one. The equivalence of (\ref{mnVolume}) and (\ref{mnVolume2}) illustrates explicitly that whereas the theory is gauge invariant under $Spin(4,1)_M$, two equivalent descriptions of the same physical quantity may have very different geometric interpretations depending on the gauge one chooses. 

\section{Concluding Remarks}
Motivated by symmetry breaking models of the Standard Model of particle physics and condensed matter systems, we have presented a quasi-dynamic symmetry breaking mechanism relating a $Spin(4,1)$ gauge theory to Einstein-Cartan gravity. The ingredients used in the construction are familiar ingredients from particle theory, namely a multiplet of Dirac spinors that are constrained to form a highly symmetric condensate. This condensate effectively breaks the symmetry, giving rise to Einstein-Cartan gravity with a topological $\theta$-term. The goal of the program is imbue the field with dynamical content and a suitable potential to {\it spontaneously} relax the multiplet to the condensate, which will presumably be an energetically or otherwise preferred ground state of the system. It should also be expected that any residual unconstrained degrees of freedom left over from the condensate of the multiplet should satisfy the dynamical content of spinorial matter, presumably given by a Dirac kinetic term in the symmetry broken action. Work towards this goal is currently in progress.

\section*{Acknowledgments}
I would like to thank Roberto Percacci for discussions and comments on a first draft of this work. This work was supported by the National Science Foundation's International Research Fellowship Program grant OISE0853116.

\appendix
\section{Spinors and the de Sitter algebra \label{AppendixA}}
Here we will review some basic properties of spinors, Clifford algebras, and their relation to the various groups used in this paper. For a more detailed discussion, see for example the appendix of \cite{Ortin}. Dirac spinors, at the most primitive level can be viewed as representations of the universal cover of the conformal group in four dimensions. By {\it Dirac spinor}, we simply mean the elements of a 4-component complex (anti-commuting) vector space subject to the inner product 
\beq
\langle \phi |\psi \rangle_D =\bar{\phi} \psi=\phi^\dagger \gamma^0 \phi \,.
\eeq
 We can then ask the question, given an arbitrary, complex and invertible $4\times 4$ matrix $g$, i.e. an element of $GL(4,\mathbb{C})$, what conditions must it satisfy to preserve this inner product? The most direct route to the answer is to choose a representation where $\g^0=i \,diag(1,1,-1,-1)$ (i.e. the Dirac representation of the Clifford algebra). The simply-connected group that preserves the inner product is then clearly $SU(2,2)$, which is isomorphic to $Spin(4,2)$, the double cover of the conformal group in 4-dimensions.

However, the question can also be easily addressed at the level of the Lie algebra, and we will gain some insight from this perspective. Clearly the group element must satisfy $-\gamma^0 g^\dagger \gamma^0 =g^{-1}$, so the Lie algebra associated with the group must satisfy
\beq
\gamma^0 \lambda^\dagger \gamma^0 =\lambda\,.
\eeq
This Lie algebra is spanned by the basis $\{\frac{1}{2}\g^{[I}\g^{J]} \,,\, \frac{1}{2} \g_5 \g^I \,,\, \frac{1}{2}\g^I \,,\, \frac{1}{4}\g_5 \}$. It can be checked that this basis spans the Lie algebra $\mathfrak{so}(4,2)$, so the unique simply connected group that preserves the Dirac inner product above is, again, $Spin(4,2)$. The conformal group has two important subgroups, the de Sitter group, $Spin(4,1)$, and the anti-de Sitter group, $Spin(3,2)$. To distinguish these groups, it is convenient to introduce two new inner products. Given the charge conjugation matrix $\mathcal{C}$ satisfying $\mathcal{C}\gamma^{I*} \mathcal{C}^{-1}=\g^I$, we define charge conjugation, and pseudo-conjugation by $\psi_c=\mathcal{C}\psi^*$ and $\psi_{c_*}=\gamma_5 \mathcal{C}\psi^*$, respectively. The Majorana inner product and pseudo-Majorana inner products are then 
\beq
\langle \phi |\psi \rangle_{M} =\bar{\phi} \psi_c \quad\quad\quad \langle \phi |\psi \rangle_{M_*} =\bar{\phi} \psi_{c_*}\,.
\eeq
The Lie algebras associated with the groups that preserve these inner products are spanned by the basis $\{\frac{1}{2}\g^{[I}\g^{J]} \,,\, \frac{i}{2}\g^{[I}\g^{J]} \,,\, \frac{1}{2}\g^I \,,\, \frac{i}{2}\g^I  \}$ for the Majorana inner product and $\{\frac{1}{2}\g^{[I}\g^{J]} \,,\, \frac{i}{2}\g^{[I}\g^{J]} \,,\, \frac{1}{2}\g_5 \g^I \,,\, \frac{i}{2}\g_5 \g^I  \}$ for the psuedo-Majorana inner product. The respective Lie algebras can then by identified as
\beq
\text{Majorana:} \ \  \mathfrak{so}(3,2)\otimes \mathbb{C} \quad\quad\quad \text{pseudo-Majorana:} \ \  \mathfrak{so}(4,1) \otimes \mathbb{C}\,.
\eeq
We can then easily distinguish the anti-de Sitter and de Sitter subgroups in a natural way. The anti-de Sitter, or rather its double cover, $Spin(3,2)$ is the unique simply connected group that preserves both the Dirac and Majorana inner products. This follows from the intersection $\mathfrak{so}(4,2) \cap (\mathfrak{so}(3,2)\otimes \mathbb{C})=\mathfrak{so}(3,2)$, and a theorem that there is a unique simply connected group associated with every semi-simple Lie algebra (see e.g. \cite{Schutz:Book}). Similarly $\mathfrak{so}(4,2) \cap (\mathfrak{so}(4,1)\otimes \mathbb{C})=\mathfrak{so}(4,1)$, so the double cover of the de Sitter group, $Spin(4,1)$, is the unique simply connected group that preserves both the Dirac and pseudo-Majorana inner products.

For a more direct identification of the de Sitter subalgebra, it is useful to extend the Clifford algebra from $Cl_{3,1}$ to $Cl_{4,1}$. It is a peculiar feature of Clifford algebras that one can generally build an odd, $N$-dimensional Clifford algebra from an $(N-1)$-dimensional even Clifford algebra without changing the dimension of the representation. For example, given the standard $4\times 4$ complex matrix representation of $Cl_{3,1}$, one enlarge the Clifford algebra by including $\g_5$ as a new element of vector part of the algebra to extend $\g^I$ where $I$ ranges from $0$ to $3$, to $\wt{\g}^A$ where $A$ ranges from $0$ to $4$. The identification is given by
\beq
\wt{\g}^I=\g^I \quad\quad\quad \wt{\g}^4=\g_5\,,
\eeq
which clearly satisfies the defining condition of the generators of $Cl_{4,1}$:
\beq
\wt{\g}^A \wt{\g}^B+\wt{\g}^B \wt{\g}^A=2\eta^{AB}\,.
\eeq
The scalar $(\bf{1})$, vector $(\wt{\g}^A)$, and bivector elements $(\wt{\g}^{[A}\wt{\g}^{B]})$ span the Clifford algebra $Cl_{4,1}$ since the analogue of the psuedo-scalar $\g_5$ in $Cl_{3,1}$ is proportional to the identity for $Cl_{4,1}$. Just as $\frac{1}{2}\g^{[I}\g^{J]}$ form a representation of the Lie algebra $\mathfrak{spin}(3,1)\simeq \mathfrak{so}(3,1)$, the bivectors $\frac{1}{2}\wt{\g}^{[A}\wt{\g}^{B]}$ form a representation of $\mathfrak{spin}(4,1)\simeq \mathfrak{so}(4,1)$. This very clearly demonstrates that Dirac spinors form a representation of $Spin(4,1)$, and it allows for a simple map between the spinor and vector representations of the group, employed in this paper by the recognition that $\bar{\psi} \tg^A \psi $ transforms as an element of the adjoint (vector) representation of $Spin(4,1)$, or equivalently the fundamental representation of $SO(4,1)$.

\bibliography{FermionCondensationBib}
 
\end{document}